# A VCSEL based Photonic Neuromorphic Processor for Event-Based Imaging Flow Cytometry Applications

M. Skontranis[1], G. Moustakas[2], A. Bogris[2], C. Mesaritakis[2]

[1])Department of Information and Communication Systems Engineering, University of the Aegean, Palama 2, 83200 Samos, Greece

[2])Department Informatics and Computer Engineering, University of West Attica, Ag. Spyridonos, 12243 Egaleo, Greece

(*Electronic mail: mskontranis@icsd.aegean.gr)

This work presents an optical neuromorphic imaging and processing cytometry system that integrates an excitable VCSEL-based time-delayed (TD) extreme learning machine with an event-based 2D camera. The proposed system is designed for the classification of Polymethyl Methacrylate (PMMA) particles of varying diameters moving at speeds between 0.01 and 0.1 m/s. The TD photonic scheme achieved a classification accuracy of 95.8% while encoding the original 2D images into a 1-bit spike stream containing a maximum of 96 spikes. Additionally, the binary representation of the synthetic frames enables a significant reduction in memory and hardware requirements, ranging from 98.4% to 99.5% and 50% to 84%, respectively. These findings demonstrate that the integration of neuromorphic computing with sensing can facilitate the development of low-power, low-latency applications optimized for resource-constrained environments.

**This work introduces a novel scheme that merges a photonic neuromorphic spiking computing scheme with a bio-inspired event-based image sensor, aiming to combine real time processing of sparse image data with increased accuracy and lightweight back-end digital processing. The neuromorphic computing scheme consists of a time-delayed spiking extreme learning machine realized through a two-section laser, biased in an excitable regime. The application targeted consists of experimental data acquired through a high-flow imaging cytometry system, aiming to classify artificial particles of different size. Results achieved demonstrate that the proposed scheme can offer comparable performance to a lightweight digital neural network (accuracy of 97.1%) but with a reduced number of parameters by a factor of 6.25. These results highlight the merits of combining neuromorphic computing with event-based sensing; paving the way for fast, low-power technologies in fields like environmental monitoring, biomedical analysis, and smart sensing.**

## I. INTRODUCTION

The last decade has been marked by the emergence of hardware systems able to implement large scale bio-inspired computational models. The rise of large scale (deep) neural networks has affected in a profound manner a wide pallet of disciplines, spanning from bioinformatics and computational chemistry to communications, natural language processing and signal analysis in general [1–3]. The enhanced processing capabilities of these systems stem from the collaborative operation of multiple nonlinear nodes, known as neurons, which communicate with one another via weighted connections known as synapses. By evolutionary design in this scheme computation coexists physically with data storage, thus fundamentally differs from the operation of classical Von Neumann digital processors. This discrepancy is the root of power inefficiency observed in large scale, state of the art neural implementation[4]. In this landscape, neuromorphic computing is a compelling alternative to digital implementations of neural networks [5,6], where Spiking Neural Network (SNNs) leverage the topologies of conventional schemes, such as feedforward and recurrent neural networks, while replacing nodes with excitable neurons. In this case, discrete, asynchronous actions potentials (spikes) encode analog values to quantities such as firing rate or time[7]. This strategy offers SNN schemes the ability to consume power solely during spike events, with minimal power consumption during idle periods [6]. Photonic technology has been proposed as a promising platform for such systems, owing to the ultrafast dynamics of excitable laser-based neurons, inherent parallelism through wavelength/time multiplexing and low loss silicon photonic synapses [6,8]. Although multiple photonic SNNs have been proposed using either excitable laser neurons [9,10], or weighted proxies [11] a pending issue is that large scale manifestations of SNNs dictate a significant number of lasers (neurons) coexisting with dense synapses (waveguides), thus raising stringent fabrication requirements (hybrid integration)[12]. On the other hand, hardware friendly implementations, relying on the reservoir computing (RC)[12] or to its spike equivalent liquid state machine (LSM)[13] has been proposed that rely on time-multiplexing laser inputs/outputs to form virtual nodes (time delayed RC or LSMs (TDRC or TD-LSM)). In this context, a single waveguide loop and neuron are utilized, thus rendering implementations more straightforward, but at the expense of a more complicated data modulation scheme (masking) and of increased processing latency[14,15]. The use of VCSELs as neurons in TD-LSMs and TDRCs can further simplify implementation due to their compact physical footprint and low power consumption[9,16–18].

In most photonic SNNs the dataset employed for their evaluation either consist of toy tasks[9,17], or they utilize tasks that are not tailored to the spiking nature of the processing module, thus cannot fully exploit their potential. In this work we aim to address this by presenting experimental/numerical results concerning a photonic TD extreme learning machine SNN scheme based on a single two section quantum well VCSEL, whereas here the dataset consists of experimental event-streams from an installed Imaging Flow Cytometry (IFC) scheme based on an event based-camera (PROPHESEE) with 1μsec temporal resolution targeting high speed particles (0.1m/s) with different diameters (12-16-20μm)[19].

## II. DATA-SET AND ARCHITECTURE

The dataset used in this work consists of experimental recordings from an IFC system targeting PMMA of three discrete diameters (12, 16, 20μm) spheres, flowing in a 100μmX100μm microfluidic channel at a speed of 0.1m/s recorded via an event-based camera. The IFC scheme comprises a LED light source emitting at 635nm, two microscope objectives, a vacuum pump and an event-based camera (PROPHESEE-EVK4). The role of the two microscope objectives was to focus/collect the light of the LED light source to the microfluidic channel whereas the vacuum pump was utilized to preserve a steady flow in the microfluidic channel whose mean velocity ranged from 0.07 to 0.1m/sec. At last, the event-based camera was used to detect the travelling particles in an asynchronous manner. In detail, EVK4 produces an event-tuple whenever a change on the intensity of a pixel was recorded. Each event was characterized by a tuple of data in the form of (X, Y, t, P) in which X and Y were the coordinates of the corresponding pixel, t is the timestamp and P is the polarity of the pixel which could be 0 or 1, depending on the decrease or increase of the intensity. EVK4 supports a temporal resolution of 1μsec per pixel whereas the spatial resolution was 640x480 pixels[19].

The generated events were used in order to generate synthetic frames, where they consisted of a 2D pixel array of accumulated events for a specific integration window ($\tau$=3msec). Therefore, each pixel in the synthetic frames is equivalent to a firing rate encoding scheme[7]. Following this step, a lightweight center of mass tracking algorithm was applied to the synthetic frames allowing the computation of the center of each particle and the cropping of the synthetic frames to 100x100 pixels[19].

The central processing unit in our case is a single VCSEL, thus in order to simulate its temporal behavior the following model was used [9]. It consists of three coupled rate equations:

$$\frac{dn_g}{dt} = \frac{I+m(t)}{q} - \frac{n_g}{\tau_g} - g_g\Gamma_g(n_g - n_{0g})N_{ph} \quad :(1)$$

$$\frac{dn_a}{dt} = -\frac{n_a}{\tau_a} - g_a\Gamma_a(n_a - n_{0a})N_{ph} \quad :(2)$$

$$\frac{dN_{ph}}{dt} = g_g\Gamma_g(n_g - n_{0g})N_{ph} + g_a\Gamma_a(n_a - n_{0a})N_{ph} - \frac{N_{ph}}{\tau_{ph}} + \beta B_r n_g^2 \quad :(3)$$

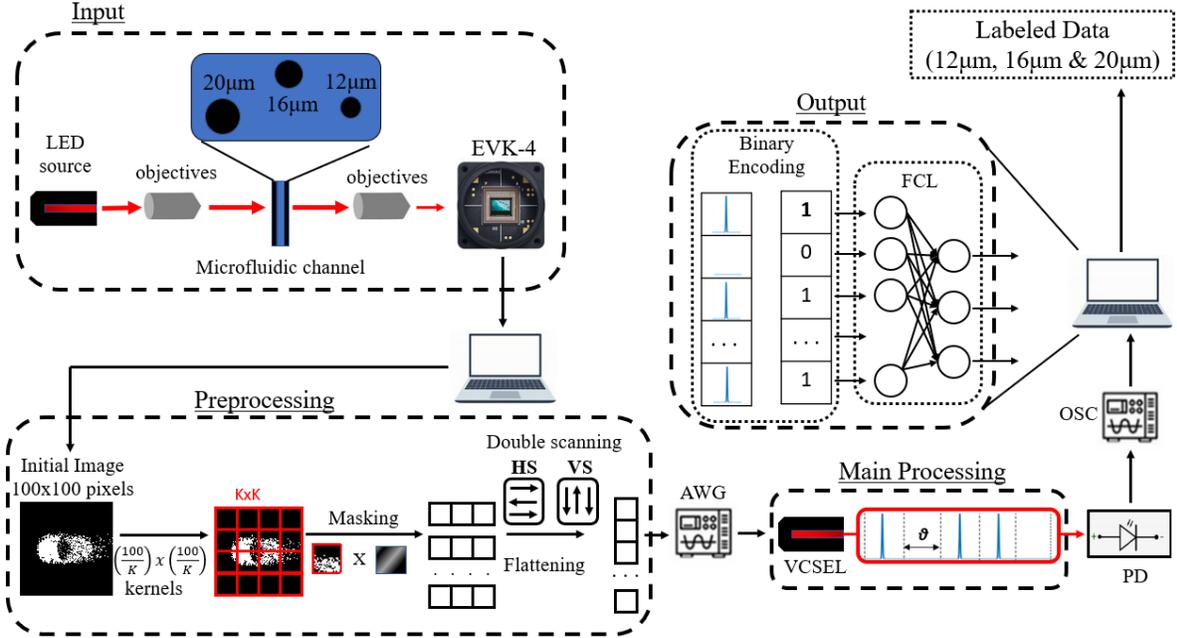

FIG. 1 Architecture of the proposed neuromorphic processor. AWG, PD and OSC stand for the Arbitrary Waveform Generator, Photodiode and Oscilloscope.

in which $n_g$ and $n_a$ are the number of electrons in the gain and absorber section area whereas $N_{ph}$ is the number of photons in the cavity. m(t) is the electrically injected signal. The subscript g and a refer to gain and absorber area, respectively. The remaining parameters are: $\tau_g$ ($\tau_a$) = 1ns (100ps) is the carrier lifetime, $\Gamma_g$ ($\Gamma_a$) = 0.06(0.05) is the confinement factor, I is the pump current, $\beta=10^{-4}$ is the spontaneous emission coefficient, $B_r = 10^{-15} m^3 s^{-1}$ is the bimolecular recombination coefficient, $g_g$ ($g_a$) = $2.9 \cdot 10^{-12} m^3 s^{-1}$ ($14.5 \cdot 10^{-12} m^3 s^{-1}$) is the differential gain and $n_{0g}$ ($n_{0a}$) = $1.1 \cdot 10^{24} m^{-3}$ ($0.89 \cdot 10^{24} m^{-3}$) is the transparency carrier density.

The processing architecture of the proposed processor is depicted in Fig.1, whereas its operation is organized in three layers designated as synthetic frame preprocessing, photonic main processing and digital output layer. The data pre-processing's task is to transform the 2D synthetic frames into a 1D sequence of analogue values similar to [17]. In particular, the initial image is organized in multiple non-overlapping kernels of KxK size. Therefore, the initial 100x100 2D array is transformed to a grid of (100/K)x(100/K) kernels (Fig.1 preprocessing). Then, each kernel is flattened (dimensions $1xK^2$) and is multiplied with a pseudo-random mask with dimensions $K^2 xL$. This process transforms the original KxK kernel to an L point vector whose values are randomized linear combinations of the original kernel.

After masking, the kernels are scanned in two different orientations: horizontal snake and vertical snake similar to [20] (see Fig.1). In the horizontal snake case, the kernels are scanned row wise, from left to right, if the row index is odd, and vice versa, if it is even, whereas a similar approach was used for the vertical snake case, where the kernels are scanned column wise, from top to the bottom, if the column index is odd, and from bottom to top, if the column index is even. This approach enhances the temporal correlation of kernels at different spatial locations of the frame. On the other hand, based on the fact that we aim for a TD approach, the increase of data points impacts processing latency[12,20]. The two scanning orientations are concatenated in a single 1 x (20000·L/K$^2$) vector. Following the TD paradigm, the elements of this vector are used to modulate the injection current of the laser-node, the modulation bandwidth is set so as to regulate the duration of each sample to a value of θ, which is known as the temporal spacing of the nodes in TD

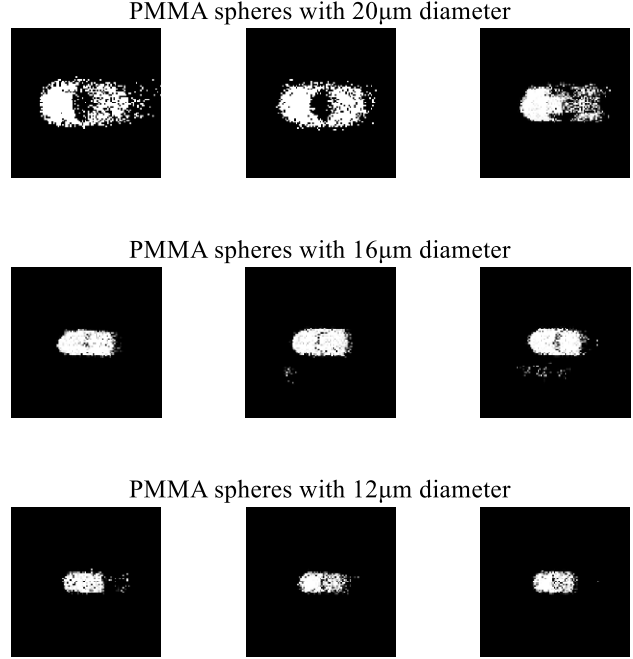

FIG. 2 Images of PMMA spheres acquired using the event-based PROPHESEE EVK4 camera. The diameters of the spheres are 20 μm (top row), 16 μm (middle row), and 12 μm (bottom row).

schemes and regulated the number of virtual nodes and their virtual interconnections[17].

The proposed photonic TD processing layer consists of a single quantum well VCSEL equipped with a saturable absorber which is biased in an excitable regime so as to be able to act as a leaky integrate and fire neuron. To be more specific, the VCSEL is forward biased (I=1.1mA), slightly below self-pulsation regime (lasing threshold $I_{th}$=1.2). Therefore, if the VCSEL's electrical injection is sufficiently perturbed (exceed a specific value, known as spiking-threshold), then an optical spike is generated. As it is obvious the role of the lasing-node is to transform the incoming analogue vector into a sparse spike train (Fig.1).

The optical output of the VCSEL is monitored by a photodiode and a real time oscilloscope (Fig.1). The digital samples recorded are compressed by encoding a 1-bit value per θ, depending on the existence or not of a spike[7]. During the final stage the binary vector is fed to a standalone digitally fully connected layer (FCL) used for frame classification, implemented through the Tensorflow module. It comprised 20000·L/K$^2$ inputs, three nodes (one for every class) and 60000·L/K$^2$ synapses. With regards to training, the cross entropy and Adam were utilized as Loss function and optimizer.

## III. RESULTS

The experimental dataset comprised 4,378 unique synthetic frames, categorized into three classes based on particle size: 1,216 frames corresponded to 20 μm spheres, 1,811 to 16 μm spheres, and 1,351 to 12 μm spheres. Typical examples of the images of the dataset are illustrated in Fig.2. Due to this class imbalance, the dataset posed a risk of introducing bias in classification performance. To address this issue, a balanced subset was constructed by randomly selecting 1,216 frames per class, resulting in a total of 3,648 images. Of these, 2,553 images (851 per class) were allocated for training, while 1,095 images (365 per class) were reserved for testing. To enhance statistical reliability, the evaluation process was repeated ten times using ten different balanced subsets of the original dataset. The final classification accuracy was determined as the average performance across these ten independent trials.

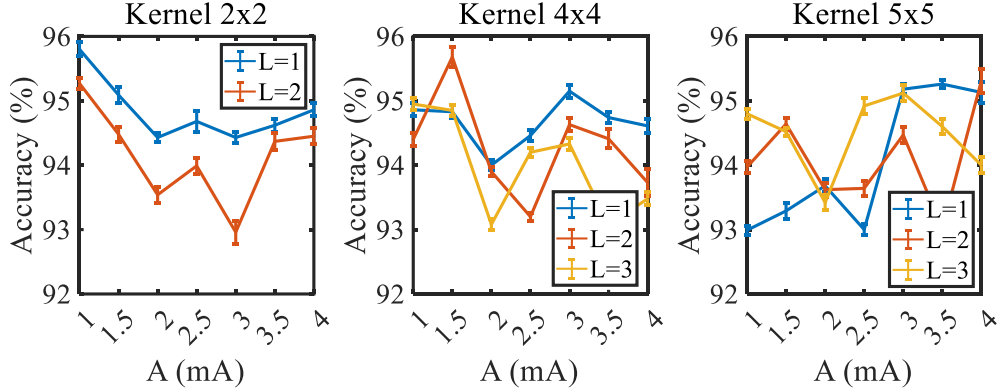

FIG. 3 Performance of the proposed processor for various kernel size, number of linear combinations per kernel (L) and amplitude of the electrical modulation (A).

The evaluation results are presented in Fig. 3. The input data were normalized in the range of (0,1) and were multiplied by amplitude of electrical modulation (A), which is a tunable parameter. Following this step, these data were used so as to provide electrically injection to the VCSEL. Each sample was injected to the processor with a sample period of 250 ps, thus setting the temporal resolution parameter (θ) to 250 ps [17]. The choice of θ influenced pixel correlation, as the VCSEL-neuron inherently correlated adjacent pixels due to its internal dynamics, including integration time and refractory period. The relaxation period of the simulated device was estimated at 1 ns, enabling spike generation at a rate of 1 GHz. Consequently, each spike carried information from four sequential pixels, defined by the ratio of the refractory period to θ (250 ps). The baseline accuracy was established at 97.1%, representing the performance obtained when the original 100 × 100 pixel synthetic frames were directly fed into a digital fully connected layer (FCL).

Three key parameters of the proposed scheme were analyzed: the kernel size (K), the number of linear combinations per kernel (L), and the amplitude of electrical modulation (A). Regarding the kernel size, three different values were considered (K = 2, 4, and 5). In the first case (K = 2), each image was represented by 5000 × L inputs. It was observed that L = 1 resulted in superior performance compared to L = 2 across all modulation amplitudes. Increasing L had a negative impact on the processor's performance, likely due to overfitting, as a larger number of inputs reduced classification capability. The highest recorded accuracy in this case was 95.8%, achieved with L = 1 and A = 1 mA.

For K = 4, each image was represented by 1250 × L inputs, effectively reducing the number of inputs by a factor of eight. The best performance recorded in this case was 95.68%, obtained with L = 2 and A = 1.5 mA. Given the limited number of inputs, the effect of L = 3 was also examined; however, as shown in Fig. 3, accuracy deteriorated in most cases. Finally, for K = 5, the number of inputs further decreased to 800 × L, leading to a 12.5-fold reduction in input size. Similar to K = 4, the highest accuracy was achieved at L = 2, but with a different modulation amplitude (A = 4 mA), reaching 95.31%. Further increasing L did not enhance performance, as L = 3 yielded a maximum accuracy of 95.12% at A = 3 mA.

The utilization of binary outputs significantly reduced both power consumption and memory requirements. The original synthetic frames exhibited a mean nonzero pixel ratio ranging from 19.12% to 27.73%, depending on the sphere class. In contrast, under the optimal configuration (K = 2, L = 1), the proposed processor generated a maximum of 96 spikes per image, with fewer than 1.92% of the inputs contributing spikes. For K = 4 and K = 5, the maximum number of spikes per image in their respective optimal cases was further reduced to 79, corresponding to 3.16% and 4.94% of the inputs generating spikes, respectively. The simplified architecture of the proposed processor led to a substantial reduction in memory and hardware demands. While a conventional fully connected neural network would require 320 Kbits and 30,000 synapses to store all inputs, the proposed scheme required only 5K bits and 15,000 synapses to achieve 95.8% accuracy, 2.5K bits and 7,500 synapses for 95.68% accuracy, and 1.6 Kbits and 4,800 synapses to reach 95.31%. Consequently, the proposed methodology effectively classified images of moving particles with a minimal accuracy loss of 1.3–1.8% while achieving a significant reduction in memory usage (98.43% to 99.5%) and hardware demands (50% to 84%).

## IV. CONCLUSION

This work presents a neuromorphic TD processor based on a single VCSEL and an event-based camera. The proposed processor unifies the disciplines of neuromorphic computing and neuromorphic sensing under a single platform. The performance of the processor is evaluated at an IFC task, classifying moving PMMA spheres based on their diameter. By exploiting the inherent advantages of both disciplines, memory demands are lowered by almost two orders of magnitude whereas hardware demands are significantly reduced by maximum of 84%. This allows the recognition of moving particles with simplified structures, paving the way towards the replacement of power hungry ANNs with energy efficient neuromorphic processors which are extremely suitable for environments where memory and energy resources are limited.

## FUNDING

This work has been partially supported by the Project QUASAR which is implemented in the framework of H.F.R.I call "Basic research Financing (Horizontal support of all Sciences)" under the National Recovery and Resilience Plan "Greece 2.0" funded by the EU – NextGenerationEU No: 016594 and EU Horizon Europe Programme PROMETHEUS under grant agreement 101070195.

## DATA AVAILABILITY STATEMENT

The data that support the findings of this study are available from the corresponding author upon reasonable request.

## DISCLOSURES

The authors declare no conflicts of interest.